\documentclass[%
preprint,
 amsmath,amssymb,
 aps,
pra,
]{revtex4-1}

\usepackage{graphicx}
\usepackage{bm}

\begin{document}

%
\title{Fe$^{\bf{15+}}$ dielectronic recombination and the effects of configuration interaction between resonances with different captured electron principal quantum number}

\author{Duck-Hee Kwon$^{1,2}$ and Daniel Wolf Savin$^{1}$}
\affiliation{$^1$Columbia Astrophysics Laboratory, Columbia University, New York, New York 10027, USA \\
$^2$Laboratory for Quantum Optics, Korea Atomic Energy Research Institute, Daejeon 305-600, Republic of Korea}



\begin{abstract}
Dielectronic recombination (DR) of Na-like Fe$^{15+}$ forming Mg-like Fe$^{14+}$ via
excitation of a $2l$ core electron has been investigated.
We find that configuration interaction (CI) between DR resonances with different captured electron principal quantum numbers
$n$ can lead to a significant reduction in resonance strengths for $n \geq 5$.
Previous theoretical work for this system has not considered this form of CI.  Including it accounts for most of the discrepancy between previous theoretical and experimental results.
\begin{description}
\item[PACS numbers]
34.80.Lx
\end{description}
\end{abstract}

\maketitle


\section{\label{sec:intro}Introduction}

Understanding the properties of astrophysical and laboratory plasmas
necessitates knowing the ionization balance of the observed or modeled sources.
This in turn depends on the underlying recombination and ionization processes.
Of particularly importance are data for the electron-ion recombination process
known as dielectronic recombination (DR) which is the dominant recombination
mechanism for most ions in atomic plasmas \cite{Bryans,Ferland}.

The DR process can be expressed as
\begin{equation}
{\rm e}^{-} + {\rm A}^{q+}_i \longleftrightarrow [{\rm A}^{(q-1)+}_j]^{**} \rightarrow [{\rm A}^{(q-1)+}_f]^{*} + \omega .
\end{equation}
DR is a two-step recombination process which begins when a free electron ${\rm e}^-$ collides with an ion of element ${\rm A}$ with charge $q+$ and in initial state $i$. The incident electron collisionally excites a core electron of the ion with principal quantum number $n_c$ and is simultaneously captured forming a system of state $j$. This process is known as dielectronic capture. We use the word ``core'' here to distinguish initially bound electrons from the captured electron. The energy of the intermediate system $[{\rm A}^{(q-1)+}_j]^{**}$ lies in the continuum and it may autoionize. DR occurs when the state $j$ radiatively decays to a state $f$ emitting a photon of angular frequency $\omega$. This reduces the total energy of the recombined system to below its ionization threshold. Conservation of energy requires that the energy of the initial free electron and unrecombined ion balance that of the intermediate
recombined system.  Thus the relative kinetic energy of the incident
electron equals the excitation energy $\Delta E$ of the core electron in
the recombined system in the presence of a captured electron plus the
binding energy $E_b$ of this captured electron in the recombined
system, i.e., $\Delta E = E_k + E_b$. Because $\Delta E$ and $E_b$ are quantized, $E_k$ is quantized, making DR a resonance process.


Here we explore a particularly nagging discrepancy between theory and experiment for the simple M-shell ion Na-like Fe$^{15+}$ forming Mg-like Fe$^{14+}$. Good agreement between experiment \cite{Linkemann} and theory has been found for Fe$^{15+}$($1s^2 2s^2 2p^6 3s$) DR via $\Delta n_c = 0$ and $1$ excitations of a $3s$ electron \cite{Altun, Gu}. For DR via $\Delta n_c = 1$ core excitation of a $2l$ electron, previous theoretical work has shown the importance of configuration interaction (CI) within a $2s^2 2p^5 3l 3l' nl''$ complex for a fixed $n$ \cite{Gor}. Including this single-$n$ CI reduced the predicted resonance structure by a factor of $2$. However, that work plus other recent work \cite{Altun, Gu}, which also consider CI only within the same $n$ complex, are still about a factor of $2$ times larger than experiment \cite{Linkemann} for resonance energies above 650 eV \cite{Gor1}. These resonances involve captured electron quantum numbers of $n \geq 5$.


In this work we investigate the cause of this discrepancy. We use the FAC (Flexible Atomic Code) \cite{Fac} which is fully relativistic and utilizes the distorted wave approximation. We have made a more complete accounting of possible autoionization and radiative decay channels than previous theoretical works. Additionally, we pay particular attention to the effect of CI between different $n$ configurations. This multi-$n$ form of CI has been neglected in previous theoretical studies for this system.

The rest of this paper is organized as follows. In Sec.~\ref{sec:theory} we review the standard theoretical approach to calculate DR, discuss the autoionization and radiative decay channels we considered for Fe$^{15+}$ DR, and outline our approach to handling CI between different $n$ complexes. We compare our theoretical results to experiment and previous theory in Sec.~\ref{results}. Lastly, we summarize our results in Sec.~\ref{sec:sum}.

\section{\label{sec:theory}Theoretical method}

\subsection{\label{sec:standard} Standard approach}
We calculated DR using the independent process, isolated resonance (IPIR) approximation \cite{Pinzola}. This method treats radiative recombination and DR separately and neglects quantum mechanical interferences between the two and between DR resonances. These interference effects have been shown to be small in general \cite{Pinzola}. The DR cross section in the IPIR approximation for a multiply-excited intermediate state $|\phi_j \rangle$ with resonance energy $E_j$ is given to lowest order in perturbation theory by \cite{Shore}


\begin{equation}
\sigma_{j}(E) = \frac{g_i}{2g_j} \frac{2 \pi^2}{k^2} \sum_f {\left \vert \frac {\langle \Phi_f | {\bf D} |  \phi_j \rangle \langle \phi_j | {\bf V} | \Psi_i \rangle} {E-E_j + i \Gamma_j/2} \right \vert}^2 .
\end{equation}
Atomic units are used here and throughout the paper unless otherwise noted. $E$ is the collision energy and $k$ is the linear momentum of incident free electron, both given in the electron-ion center-of-mass frame, $g_i$ and $g_j$ are statistical weights, $|\Psi_i \rangle$ is an initial recombining state which includes the incident free electron, and $|\Phi_f \rangle$ is a final bound state. The incident free electron is not correlated with the target ion. ${\bf D}$ is the dipole radiation field interaction
\begin{equation}
{\bf D} = \left(\frac{4\omega^3}{3 c^3}\right)^{1/2\ }\sum_{s=1}^{N+1} {\bf r}_s,
\end{equation}
where $c$ is the light velocity, $N$ is the number of bound electrons before dielectronic capture and ${\bf r}_s$ is the position vector of electron $s$ from the nucleus. ${\bf V}$ is the electrostatic interaction between the $N$ initially bound electrons and the continuum $N+1$ electron
\begin{equation}
{\bf V} = \sum_{s=1}^N \frac{1}{|{\bf r}_s - {\bf r}_{N+1}|}.
\end{equation}
The total resonance width $\Gamma_j$ is given by
\begin{equation}
\Gamma_j = \sum_k A^a_{jk} + \sum_f A^r_{jf} .
\end{equation}
$A^a_{jk}$ is the autoionization rate from $j$ to any state $k$ of A$^{q+}$ and can be expressed as
\begin{equation}
A^a_{jk} = 2\pi {\vert  \langle \Psi_k | {\bf V} | \phi_j \rangle \vert}^2.
\end{equation}
$A^r_{jf}$ is the radiative decay rate from $j$ to $f$ which can be written as
\begin{equation}
A^r_{jf} = {\vert  \langle \Phi_f | {\bf D} | \phi_j \rangle \vert}^2  .
\end{equation}
The energy integrated cross section (i.e., resonance strength) of state $j$ is given by \cite{Kilgus}
\begin{equation}
\hat{\sigma}_j = \frac{\pi^2}{E_j} \frac{g_j}{2g_i} \frac{A^a_{ji} \sum_f A^r_{jf} }{\sum_k A^a_{jk} + \sum_f A^r_{jf}} , \label{resonancestrength}
\end{equation}
in the approximation $\Gamma_j \ll E_j$. The resonance strength can be re-written as the product of the dielectronic capture (DC) strength
\begin{equation}
S_{\rm DC} = \frac{\pi^2}{E_j} \frac{g_j}{2g_i} A^a_{ji} ,
\end{equation}
which is related to the autoionization rate through detailed balance,
and the branching ratio
\begin{equation}
B_j = \frac{\sum_f A^r_{jf}}{\sum_k A^a_{jk} + \sum_f A^r_{jf}}.
\end{equation}

\subsection{\label{sec:channe}DR channels of Fe$^{\bf 15+}$}

For Fe$^{15+}$ DR via $\Delta n_c = 1$ core excitation of a $2l$ electron, we considered the autoionization and radiative decay channels
\begin{widetext}
\begin{eqnarray}
{\rm e}^- + 2s^2 2p^6 3s~^2S_{1/2} \rightarrow 2l^7 3l' 3l'' nl''' &\rightarrow&
\begin{Bmatrix} 2l^8 3l' \\ 2l^8 nl''' \\ 2l^7 3l' 3l'' \end{Bmatrix} + {\rm e}^{-} \\
&\rightarrow& \begin{Bmatrix} 2l^8 3l' 3l'' \\ 2l^8 3l' nl'''  \end{Bmatrix} + \omega ,
\end{eqnarray}
\end{widetext}
where $l \leq 1$, $l'$ and $l''$ $ \leq 2$, and $l''' \leq 5$. This includes the $2l^8 3l' 3l''$ radiative decay channel which was not considered by \cite{Gu} and the $2l^7 3l' 3l''$ autoionization channel which was not included by \cite{Gor}. We also considered CI for all possible $2l^7 3l' 3l''$ core configurations. Thus, unlike the previous theoretical work,  $2s \rightarrow 3l$ promotions are included.

For $n > 6$, the $2l^8 3l' nl'''$ configuration lies in the continuum and radiative decays to autoionizing levels are possible.  These can then autoionize or radiatively stabilize via

\begin{eqnarray}
2l^8 3l' nl''' &\rightarrow& \begin{Bmatrix} 2l^8 3l'' \end{Bmatrix} + \rm{e}^{-} \\
                     &\rightarrow& \begin{Bmatrix} 2l^8 3l'' nl''' \\ 2l^8 3l' 3l'' \\ 2l^8 3l' n' l'' \end{Bmatrix} + \omega .
\end{eqnarray}
The branching ratio for these radiative {\it decays} to {\it autoionizing} levels followed by radiative {\it cascades} (DAC) can be given by \cite{Behar}

\begin{equation}
B_j = \frac{\sum_{t} {A^r_{jt}} + \sum_{t'} A^r_{jt'} B_{t'}}{\sum_k A^a_{jk} + \sum_f A^r_{jf}} ,
\end{equation}
where the final states $t$ and $t'$ are below and above the ionization threshold, respectively. $B_{t'}$ is the branching ratio for radiative stabilization of $t'$ and can be determined by evaluating $B_j$ iteratively.

\subsection{\label{sec:CI}Configuration interaction between different \boldmath{$n$} resonance complexes}

We performed a large scale CI calculation between all $2l^7 3l' 3l'' nl'''$ complexes from $n = 3$ to $14$. This allows us to consider CI between resonances with different captured electron principal quantum numbers. A large orbital sensitivity of DR to the choice of the initial radial wave function has been reported in Mg$^{2+}$ \cite{Orb}. For Fe$^{15+}$ this sensitivity is expected to be insignificant as a result of the high $q$ of the ion. We explicitly investigated here the effects of optimizing radial wave functions on the $2l^7
3l'3l''$ and $2l^8 3l'$ configurations of the recombining ion as well as on the
$2l^7 3l'3l''3l'''$ and $2l^8 3l' 3l''$ configurations of the recombined ion.
Only small differences in resonance strengths and energies were
seen. In the end, radial wave functions were optimized on the $2l^8 3l' 3l''$ configuration as that gave best agreement
with the experimental results.

The $j$ CI mixed state $\bar{\phi}^n_j$ for an $n$ complex can be expanded in the $j'$ unmixed basis $\phi^{n'}_{j'}$ of an $n'$ complex using
\begin{equation}
\bar{\phi}^n_j = \sum_{n'}\sum_{j'}  c_{n'j'} \phi^{n'}_{j'} , \label{expansion}
\end{equation}
where $c_{n'j'}$ denotes the mixing coefficient for the $\phi^{n'}_{j'}$ basis.
We calculated autoionization and radiative decay rates from the wave functions obtained using this CI mixing. Past studies have not considered CI mixing between different $n$ complexes. In those studies autoionization channels of the form $2l^7 3l' 3l'' nl''' \rightarrow 2l^8 nl''' + \rm{e}^- $ and radiative decay channels of the form $2l^7 3l' 3l'' nl''' \rightarrow 2l^8 3l' nl''' + \omega$ were possible only between the states of same $n$. However, taking into account CI mixing between different $n$ resonance complexes allows for additional autoionization and radiative decay channels.
\begin{figure*}[t]
\includegraphics[width=17.0cm]{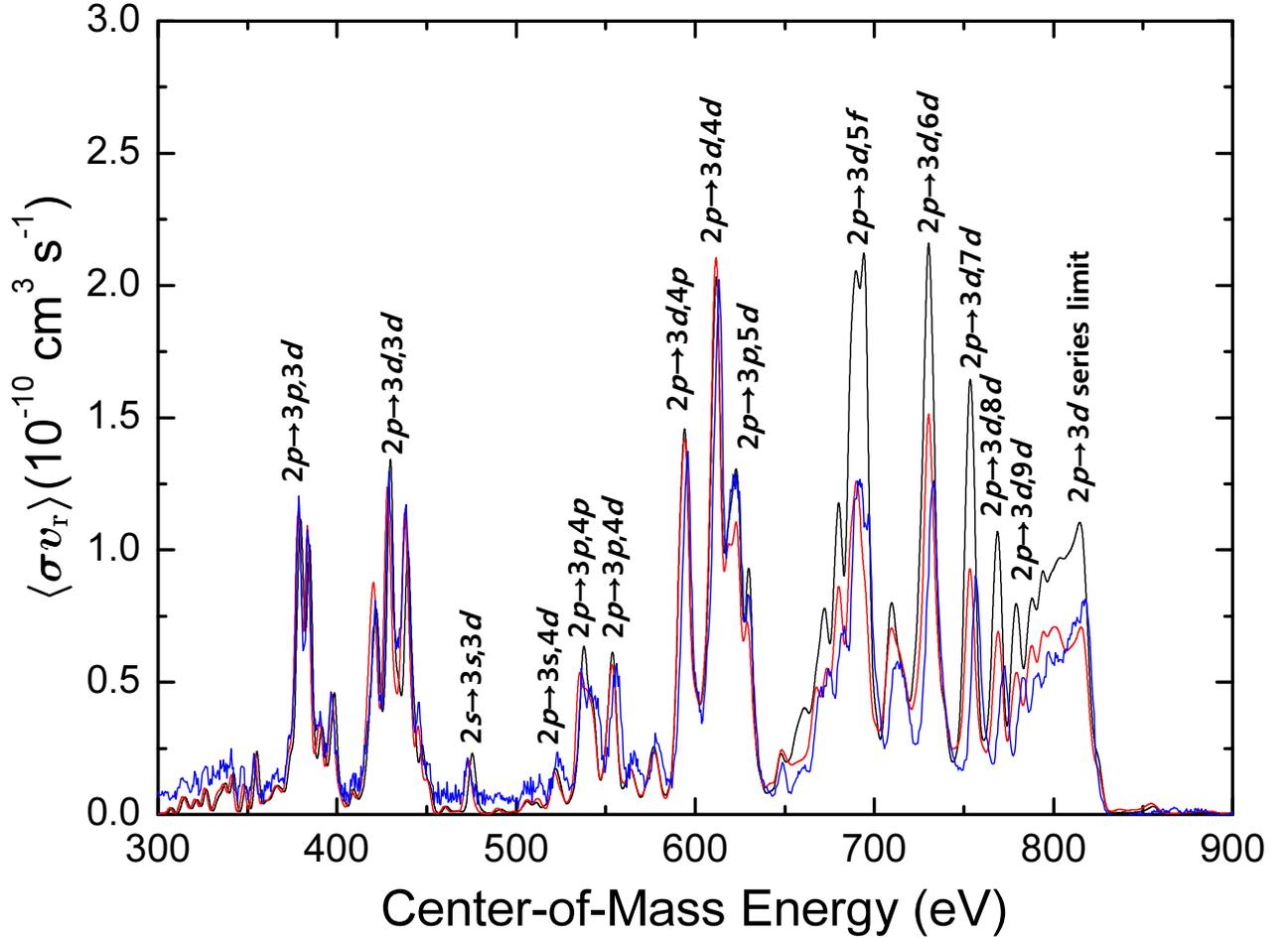}
\caption{\label{fig:DRstructure} DR resonance structure of Fe$^{15+}$ via $\Delta n_c = 1$ core excitation of a $2l$ electron. The blue line shows the experimental results of \cite{Linkemann}. The black line shows our results including CI only within the same $n$ resonance complex. The red line shows our results including CI between different $n$ resonance complexes for $3 \le n \le 14$. See text for details.}
\end{figure*}

\section{\label{results} Results}
\subsection{\label{Exp} Experiment}

Theoretical studies of Fe$^{15+}$ DR have been aided greatly by the merged-beams experimental results of \cite{Linkemann} shown in Fig.~\ref{fig:DRstructure}. The measured data represent the DR cross section $\sigma$ times the relative collision velocity $v_{\rm r}$ convolved with the experimental

\begin{table*}[t]
\caption{\label{tab:DRstrength} DR resonance energy and strength for the strongest level in each $n$ complex for single-$n$ CI. Also listed are the corresponding data for multi-$n$ CI where $3 \le n \le 14$. $J$ denotes the total angular momentum of each level. For the level description, relativistically closed shells with $J = 0$, such as $2s^2$ and $2p^4_{3/2}$, are omitted for brevity.
}
\begin{ruledtabular}
\begin{tabular}{lccccc}
            &   &\multicolumn{2}{c}{Resonance energy (eV)}&\multicolumn{2}{c}{Resonance Strength ($10^{-19}$ cm$^2$ eV)} \\
            \cline{2-4} \cline{5-6}
         Level & $J$ & Single-$n$ CI & Multi-$n$ CI & ~~~~~Single-$n$ CI & Multi-$n$ CI \\
\hline
$[(2p_{1/2}3s)_1 3p_{3/2}]_{1/2} 3d_{5/2}$ & 3  & 385.17 & 383.80 & ~~~~~~1.53 & 1.54 \\
$[(2p_{1/2}3s)_1 3d_{3/2}]_{1/2} 4d_{5/2}$ & 3  & 612.54 & 612.51 & ~~~~~~2.03 & 1.94 \\
$[(2p_{1/2}3s)_1 3d_{3/2}]_{1/2} 5f_{7/2}$ & 4  & 694.94 & 695.20 & ~~~~~~1.64 & 0.07 \\
$[(2p_{1/2}3s)_1 3d_{3/2}]_{1/2} 6d_{5/2}$ & 3  & 730.04 & 730.17 & ~~~~~~1.33 & 0.22 \\
$[(2p_{1/2}3s)_1 3d_{3/2}]_{1/2} 7d_{5/2}$ & 3  & 754.21 & 754.16 & ~~~~~~0.96 & 0.47 \\
$[(2p_{1/2}3s)_1 3d_{3/2}]_{1/2} 8d_{5/2}$ & 3  & 769.73 & 769.75 & ~~~~~~0.68 & 0.40 \\
$[(2p_{1/2}3s)_1 3d_{3/2}]_{1/2} 9d_{5/2}$ & 3  & 780.30 & 780.30 & ~~~~~~0.53 & 0.32 \\
$[(2p_{1/2}3s)_1 3d_{3/2}]_{1/2} 10d_{5/2}$ & 3  & 787.81 & 787.83 & ~~~~~~0.41 & 0.25 \\
$[(2p_{1/2}3s)_1 3d_{3/2}]_{1/2} 11d_{5/2}$ & 3  & 793.35 & 793.35 & ~~~~~~0.32 & 0.21 \\
$[(2p_{1/2}3s)_1 3d_{3/2}]_{1/2} 12d_{5/2}$ & 3  & 797.54 & 797.55 & ~~~~~~0.25 & 0.13 \\
$[(2p_{1/2}3s)_1 3d_{3/2}]_{1/2} 13d_{5/2}$ & 3  & 800.08 & 800.81 & ~~~~~~0.20 & 0.13 \\
$[(2p_{1/2}3s)_1 3d_{3/2}]_{1/2} 14d_{5/2}$ & 3  & 803.37 & 803.41 & ~~~~~~0.17 & 0.07 \\
\end{tabular}
\end{ruledtabular}
\end{table*}
energy distribution yielding a rate coefficient $\langle \sigma v_{\rm r} \rangle$ \cite{Kilgus}. The energy distribution is described by a flattened Maxwellian with a temperature of $k_B T_{\|} = 2.4$~meV along the beams and a temperature of $k_B T_{\bot} = 0.1$~eV perpendicular to the beams. Field ionization in the experiment limits the measured data to $n \lesssim 86$.

\subsection{\label{CI1} CI within the same \boldmath{$n$} complex}

We performed explicit calculations of autoionization and radiative decay rates up to $n = 14$ and extrapolated for $n$ from $15$ to the experimental cutoff of $86$. A simple hydrogenic scaling law was used for the resonance energies, the autoionization rates, and the radiative decay rates of the captured electron for $n \ge 15$. The radiative decay rate of the core electron was set to the $n = 14$ value for all $n \geq 15$. The calculated DR strengths were multiplied by $v_{\rm r}$ and convolved with the experimental energy distribution of \cite{Linkemann}. The results are shown in Fig.~\ref{fig:DRstructure}. In the figure we have also labeled some of the strong resonances based on the results of our calculations. Table~\ref{tab:DRstrength} lists the calculated resonance energies and strengths for the strongest resonance level in each $n$ complex for $3 \leq n \leq 14$.
Including the $2s \rightarrow 3l$ promotion channel gives improved agreement between theory and experiment in the collision energy range of 400--500 eV. The resonances between 400--450 eV are in the better agreement with the experiment compared to the previous FAC results \cite{Gu}. Also the resonance at $\sim470$ eV does not appear unless this excitation channel is included. However including CI only within the same $n$ complex does not remove the large discrepancy between theory and experiment for collision energies over 650 eV.


\subsection{\label{CI2} CI between different {\boldmath$n$} complexes}

Explicit calculations for autoionization and radiative decay rates were again carried out to $n = 14$. For higher $n$, the extrapolation described in Sec.~\ref{CI1} was performed. The convolved results are shown in Fig.~\ref{fig:DRstructure}. Resonance strengths and energies are reported in Table~\ref{tab:DRstrength} for the selected resonances described in Sec.~\ref{CI1}. Figure~\ref{fig:DRstructure} shows that above $\sim 650$~eV multi-$n$ CI dramatically reduces the theoretical results compared to single-$n$ CI. This reduction brings theory into very good agreement with experiment. The previous factor of 2 differences have been reduced to the level of tens of percent. The remaining differences near the series limit may be due to field ionization effects in the experiment as described by \cite{Schipper}, computational resources having limited the multi-$n$ CI calculations to $n \le 14$, or some combination thereof.

\begin{figure}[h]
\includegraphics[width=8.5cm]{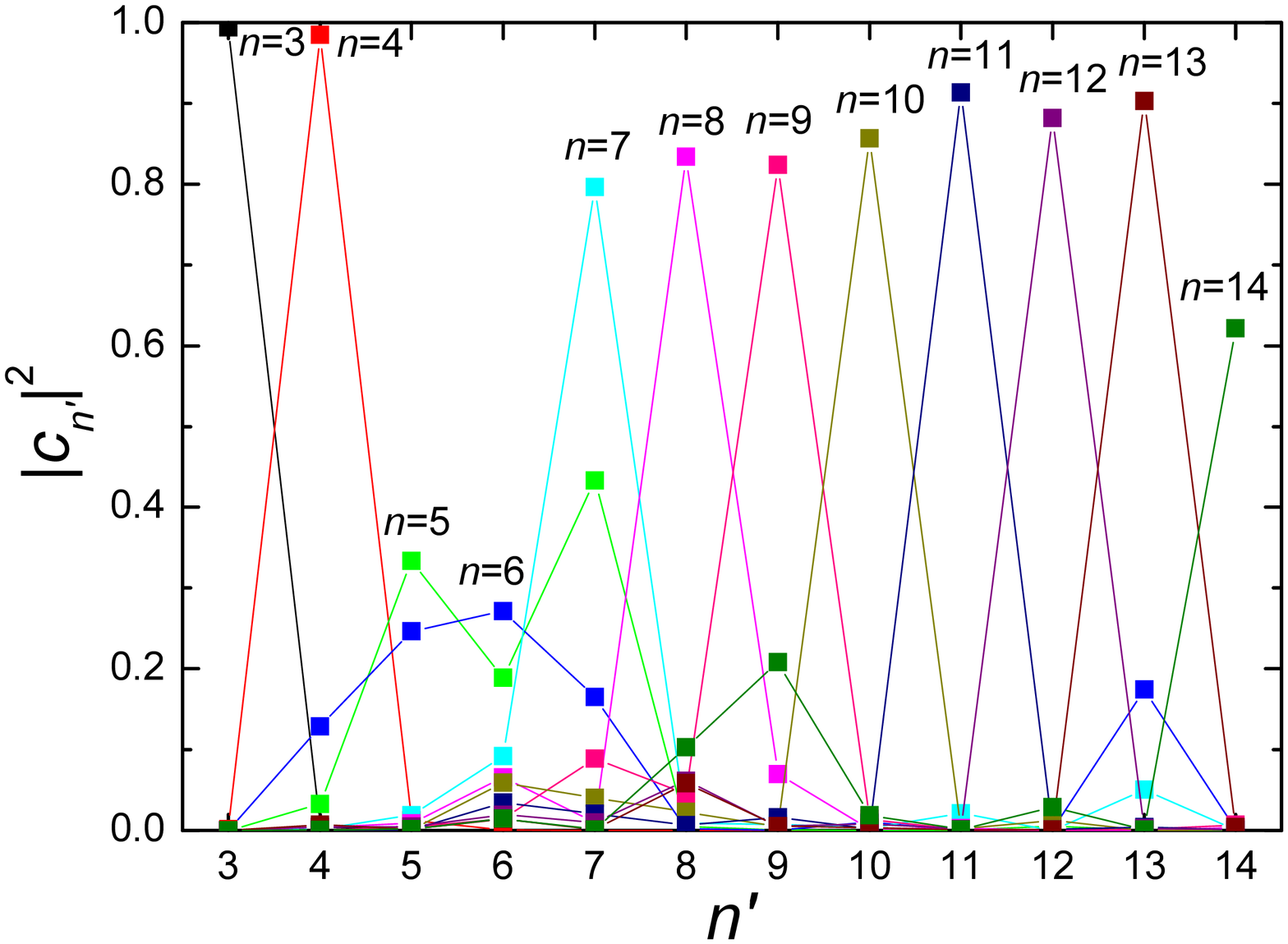}
\caption{\label{fig:CIcoeff}Mixing factor $|c_{n'}|^2$ as a function of the mixing complex $n'$. Results are plotted for the resonances listed in Table~\ref{tab:DRstrength}. Each curve is labeled by the initial $n$ configuration before mixing between $n'$ configurations is included.}
\end{figure}

A general sense for the importance of multi-$n$ CI for $n \ge 5$ can be gained by looking at the mixing factors for the resonances listed in Table~\ref{tab:DRstrength}. The mixing factor is given by
\begin{equation}
|c_{n'}|^2 = \sum_{j'} |c_{n'j'}|^2 , \label{mixfactor}
\end{equation}
where the summation is over the $j'$ basis states in the $n'$ complex. The mixing occurs between levels with the same parity, symmetry, and angular momentum. The mixing factors are plotted in Fig.~\ref{fig:CIcoeff}. One sees that the $n = 3$ and $4$ resonances of Table~\ref{tab:DRstrength} are largely unmixed with other $n'$ complexes but the $n \ge 5$ resonances can be strongly mixed. In particular the $n=5$, 6, and 14 resonances are very strongly mixed with other $n'$ complexes.

To gain a more quantitative understanding on how multi-$n$ CI can affect the predicted resonance strengths, it is helpful now to re-write Eq.~(\ref{resonancestrength}) using the expansion basis of Eq.~(\ref{expansion}) which gives
\begin{widetext}
\begin{equation}
\hat{\sigma}_j = \frac{\pi^2}{E_j} \frac{g_j}{2g_i} \frac{ \sum_{n'}\sum_{j'} |c_{n'j'}|^2  A^a_{n'j'i} \sum_{n'}\sum_{j'} (|c_{n'j'}|^2 \sum_f A^r_{n'j'f})}
{ \sum_{n'}\sum_{j'} |c_{n'j'}|^2 ( \sum_k A^a_{n'j'k} + \sum_f A^r_{n'j'f} ) } .
\end{equation}
\end{widetext}
Here $A^a_{n'j'i}$ is the autoionization rate from the unmixed basis state $\phi^{n'}_{j'}$ to an initial state $i$ and is given by
\begin{equation}
A^a_{n'j'i} = 2\pi {\vert  \langle \Psi_{i} | {\bf V} | \phi^{n'}_{j'} \rangle \vert}^2. \label{auto}
\end{equation}
$A^a_{n'j'k}$ is given by Eq.~(\ref{auto}) but changing $i \rightarrow k$. $A^r_{n'j'f}$ is the radiative decay rate from the $\phi^{n'}_{j'}$ to a state $f$ and is given by
\begin{equation}
A^r_{n'j'f} = {\vert  \langle \Phi_f | {\bf D} | \phi^{n'}_{j'} \rangle \vert}^2.
\end{equation}
The coupling (i.e., interference) terms between different basis such as $\langle \Psi_k | {\bf V} | \phi^{n'}_{j'} \rangle \langle \phi^{n''}_{j''} | {\bf V} | \Psi_k \rangle $ and $\langle \Phi_f | {\bf D} | \phi^{n'}_{j'} \rangle \langle \phi^{n''}_{j''} | {\bf D} | \Phi_f \rangle $ have been neglected just as in the IPIR approximation. The dielectronic capture strength for the CI mixing can be re-expressed as
\begin{equation}
S_{\rm{DC}} = \frac{\pi^2}{E_j} \frac{g_j}{2g_i} \sum_{n'}\sum_{j'} |c_{n'j'}|^2  A^a_{n'j'i} ,\label{capture}
\end{equation}
and the branching ratio for the CI mixing is given by
\begin{equation}
B_j = \frac{ \sum_{n'}\sum_{j'} |c_{n'j'}|^2 (\sum_f A^r_{n'j'f})}
{ \sum_{n'}\sum_{j'} |c_{n'j'}|^2 ( \sum_k A^a_{n'j'k} + \sum_f A^r_{n'j'f} ) }. \label{branch}
\end{equation}

\begin{table*}[h]
\caption{\label{tab:mixlevel} Mixing basis level distribution for the $[(2p_{1/2}3s)_1 3d_{3/2}]_{1/2} 6d_{5/2}$ resonance listed in Table~\ref{tab:DRstrength}. For the level description, relativistically closed shells with $J = 0$, such as $2s^2$, $2p^2_{1/2}$ and $2p^4_{3/2}$, are omitted for brevity. The square of the mixing coefficient as defined in Eq.~(\ref{expansion}) is given in percent. $A^a_{n'j'i}$ is the autoionization rate from $j'$ to $i$ where $i$ is the initial state $2s^2 2p^6 3s~^2S_{1/2}$ of recombining ion,  $\sum_k A^a_{n'j'k} + \sum_f A^r_{n'j'f}$ is the total autoionization and radiative decay rate of $j'$, and $B_{j'}$ is the branching ratio of $j'$. Only basis levels where $|c_{n'j'}|^2 > 2\%$ are listed. A total of 10298 basis levels were included for this $n = 6$ resonance.
}
\begin{ruledtabular}
\begin{tabular}{lcccc}
$j'$ basis level & $|c_{j'n'}|^2$   & $A^a_{j'n'i}$ & $\sum_k A^a_{j'n'k} + \sum_f A^r_{j'n'f}$ & $B_{j'}$ \\
                 &   (\%)       &   (s$^{-1}$)                      & (s$^{-1}$) &  \\
\hline
$[(2p_{1/2}3s)_1 3d_{3/2} ]_{1/2} 6d_{5/2}$ & 14.4 & $2.64 \times 10^{13}$ & $5.77 \times 10^{13}$ & 0.423 \\
$[(2p^3_{3/2}(J=3/2)3s)_2 3p_{3/2}]_{5/2} 13h_{11/2}$ & 8.4 & $3.17 \times 10^{7}$ & $5.24 \times 10^{11}$ & 0.032 \\
$[(2p^3_{3/2}(J=3/2)3s)_2 3d_{5/2} ]_{1/2} 6d_{5/2}$ & 6.8 & $2.70 \times 10^{12}$ & $ 8.46 \times 10^{13}$ & 0.051 \\
$[(2p^3_{3/2}(J=3/2)3s)_1 3p_{3/2}]_{5/2} 13h_{11/2}$ & 5.8 & $4.38 \times 10^{7}$ & $8.24 \times 10^{11}$ & 0.012 \\
$[(2p_{1/2} 3p_{3/2})_2 3d_{3/2}]_{3/2} 5f_{5/2}$ & 5.8 & $1.40 \times 10^{10}$ & $3.49 \times 10^{13}$ & 0.513 \\
$(2p^3_{3/2}(J=3/2) 3p^2_{3/2}(J=2))_{3/2} 7g_{9/2}$ &4.8 & $1.40 \times 10^{10}$ & $1.43 \times 10^{13}$ & 0.003 \\
$[(2s 3p_{1/2})_1 3d_{5/2}]_{5/2} 4s$ &4.4& $4.22 \times 10^{10}$ & $6.97 \times 10^{12}$ & 0.348 \\
$[(2p^3_{3/2}(J=3/2) 3s)_{1} 3d_{5/2}]_{5/2} 7s$ & 2.6 & $4.22 \times 10^{10}$ & $1.33 \times 10^{13}$ & 0.003 \\
$[(2p_{1/2} 3p_{3/2})_1 3d_{3/2}]_{3/2} 5f_{5/2}$ & 2.3 & $ 5.64 \times 10^{8}$ & $2.08 \times 10^{12}$ & 0.131 \\
$[(2s 3p_{1/2})_1 3d_{3/2}]_{5/2} 4s$ &2.3& $1.00 \times 10^{11}$ & $3.17 \times 10^{13}$ & 0.022 \\
\end{tabular}
\end{ruledtabular}
\end{table*}

Now, taking the $[(2p_{1/2}3s)_1 3d_{3/2} ]_{1/2} 6d_{5/2}$ resonance level of the $n=6$ complex listed in Table~\ref{tab:DRstrength} as an example, we find that it mixes primarily with the basis levels listed in Table~\ref{tab:mixlevel}. Note that the autoionization rate $A^a_{n'j'i}$ from the $[(2p_{1/2}3s)_1 3d_{3/2} ]_{1/2} 6d_{5/2}$ level to $i$ is over a factor of 10 larger than the autionization rates from the other listed basis levels to $i$. This leads to a reduction in $S_{\rm DC}$ by a factor of 6.5 when the values listed in Table~\ref{tab:mixlevel} are used in Eq.~(\ref{capture}), compared to what is calculated for the single-$n$ CI case.

\begin{figure}[h]
\includegraphics[width=8.5cm]{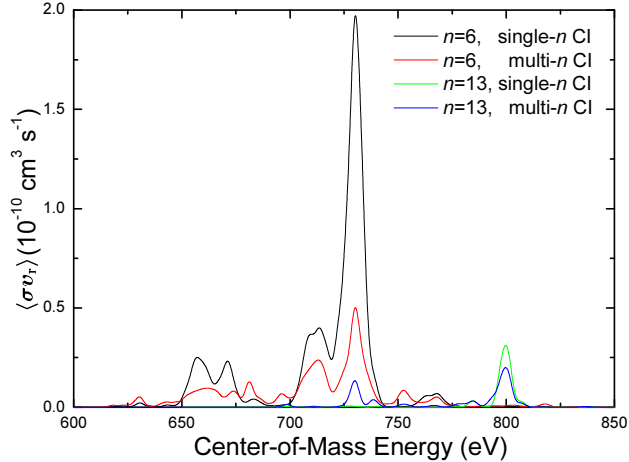}
\caption{\label{fig:reso} DR resonance structure of the $n = 6$ and $n = 13$ resonances for single-$n$ CI and for multi-$n$ CI.}
\end{figure}

The branching ratios for the listed levels which mix with the selected $n=6$ resonance are all similar or smaller in value to that for this specific level. The resulting total branching ratio $B_j$ given by Eq.~(\ref{branch}) is reduced by a factor of 1.3 from the single-$n$ CI value. Combining the multi-$n$ values for $S_{\rm DC}$ and $B_j$, we find a reduction for the resonance strength of the $[(2p_{1/2}3s)_1 3d_{3/2} ]_{1/2} 6d_{5/2}$ level by a factor of 8.5, compared to the single-$n$ CI results. This estimate agrees reasonably well with the factor of 6.5 reduction from the more complete calculation, as can be seen in Table~\ref{tab:DRstrength}. The convolved DR resonance strengths for all $n=6$ resonances are displayed in Fig.~\ref{fig:reso}. The reduction of the strong $n = 6$ resonances can be clearly seen. The reduction of the resonance strength for the other resonances listed in Table~\ref{tab:DRstrength} can be similarly explained by multi-$n$ CI mixing. In short, the DR resonance strength for strong resonances can be decreased by CI mixing with weak resonances.

On the other hand, the DR resonance strengths for some weak resonances can be increased by CI mixing with strong resonances of different $n'$ complexes. For example, the $n=13$ resonance strengths around 730~eV are largely enhanced by CI with the strong $n = 6$ resonances at this energy, as is shown in Fig.~\ref{fig:reso}.

\section{\label{sec:sum}Summary}

We have demonstrated the importance of CI between resonances with different captured electron principal quantum numbers $n$ for DR of Na-like Fe$^{15+}$ forming Mg-like  Fe$^{14+}$ via $\Delta n_{\rm c} = 1$ core excitation of a $2l$ electron. Multi-$n$ CI significantly reduces the theoretical resonance strengths for capture into $n \ge 5$ levels which overlap in energy with other many different $n$ levels. This brings theory into very good agreement with experiment and removes a previously existing discrepancy between the two. The $n=4$ levels are largely unaffected by multi-$n$ CI because the energy separation between the $n=4$ resonances and the interacting
higher $n$ resonances is large enough to render the multi-$n$ CI unimportant.
Such is not the case for the energy separation of the $n \ge 5$ resonances
and those that they interact with, particularly for $n = 5$, 6, and 14. Additionally we have shown the importance of DR via $2s \to 3l$ core promotions.

\section*{Acknowledgements}

We thank M. F. Gu for providing support and guidance in using his Flexible Atomic Code (FAC) programs. We would like to thank N. R. Badnell and T. W. Gorczyca for stimulating discussion about this work and A. M\"{u}ller for providing the experimental data. This work was supported by the NASA Astronomy and Physics Research and Analysis (APRA) Program and the NASA Solar and Heliospheric Physics Supporting Research and Technology Program.

\bibliography{apssamp}

\end{document}